\documentclass{aa}  

\usepackage{graphicx}
\usepackage{txfonts}
\usepackage{xcolor}

\begin{document} 

   \title{Local mixing length theory with compositional effects:\ First application to asymptotic giant branch evolution}
   \author{M.M. Ocampo,
          \inst{1,2}
          M.M. Miller Bertolami\inst{1,2}, A.H. C\'orsico\inst{1,2} \and L.G. Althaus\inst{1,2} 
          }

   \institute{Instituto de Astrof\'isica de La Plata, CONICET-UNLP, Argentina\\
         \and
             Facultad de Ciencias Astron\'omicas y Geof\'isicas, Universidad Nacional de La Plata, Argentina\\
             \email{mocampo@fcaglp.unlp.edu.ar}
             }


 
  \abstract
   {During the evolution of stars on the asymptotic giant branch (AGB), thermal pulses lead to the formation of strongly stratified layers in the outer regions of the CO core, which might lead to inversions in the chemical gradient. Such inversions would produce instabilities beyond the ones predicted by the Schwarzschild criterion and the standard use of mixing length theory (MLT). }
   {We aim to broaden the scope of MLT in a simple way to consider Rayleigh-Taylor and thermohaline instabilities. We also want to explore the impact of such instabilities during the AGB phase and in the pulsational properties of variable PG1159 (GW Vir) stars.}
   {We used a set of MLT equations that consider the impact of the background chemical gradients. This extension of MLT is referred to in this work as MLT$\sharp$,  to make a distinction between both prescriptions. We  applied MLT$\sharp$   in tandem with the more general Ledoux instability criterion. We computed the evolution in the AGB phase and compared the chemical profiles resulting from MLT, MLT$\sharp$ and the double diffusive GNA theory. We continued the evolution through a post-AGB thermal pulse and performed a  pulsational analysis of the resultant GW Vir models to asses $g$-mode pulsation periods. Finally, we tested our results with pulsation properties of known GW Vir stars derived from recent observations. }
   {We find that the much simpler MLT$\sharp$ set of equations closely reproduces the results from the GNA theory. As such, MLT$\sharp$ offers a simple way to include chemically driven convection in stellar evolution computations. Stellar evolution simulations show that Rayleigh-Taylor and thermohaline instabilities can play an important role during the TP-AGB. We obtained significantly different chemical profiles using a standard MLT approach compared to those resulting from our MLT$\sharp$ and GNA computations. Our adiabatic pulsational analysis shows that these differences in the chemical stratification leave clear mode-trapping signatures in the pulsation spectrum of the GW Vir models. Even though the comparison with current available GW Vir pulsation periods does not conclusively favor any of the three models of mixing explored in this work, our analysis demonstrates that asteroseismology of GW Vir stars has significant potential for probing chemically driven mixing processes during the TP-AGB phase.}
  {}


    \authorrunning{Ocampo et. al.}
    \titlerunning{}
   \maketitle

%

\section{Introduction}
The turbulent transport of heat, angular momentum, and chemical species stands among the most important unsolved problems in the modeling of stellar interiors \citep[e.g.,][]{2015ApJ...809...30A,2017RSOS....470192S,2021FrASS...7...95X,2022ApJ...926..169A,2023A&A...680A.101S}.
Convective energy transport, as well as mixing processes in stellar astrophysics, are usually described with the help of the so-called mixing length theory (MLT; \citealt{1925Prandtl,1932Biermann,1953ZA.....32..135V,1958ZA.....46..108B}). This theory, together with the so-called Schwarzschild instability criterion \citep{1906WisGo.195...41S}, provides a simple algorithm to treat convection locally at any given point in the star. Despite its simplicity, it shows a good agreement with observations. In its standard and most adopted form, MLT does not take into account the impact of chemical gradients on the development of convective motions \citep{2020mdps.conf...13G}, nor does it describe the more subtle double-diffusive mixing processes, such as thermohaline mixing \citep{1980A&A....91..175K} and semiconvection \citep{1966PASJ...18..374K,1983A&A...126..207L}.

\cite{1993ApJ...407..284G}  and \cite{1996MNRAS.283.1165G}, henceforth referred to as GNA, developed a theory that considers the Rayleigh-Taylor (RT) instabilities as well as the double-diffusive mixing processes that can develop in the presence of chemical gradients. This is a nonlocal description of convection equipped with its own stability conditions and successfully reproduces most known results in stellar convection \citep{2011A&A...533A.139W}. However, GNA does not consider the possible electronic degeneration and nonideal effects in the stellar medium and in evolved stars, where the deep interior becomes degenerate, this description can be insufficient as well. Moreover, the mathematics involved in GNA are far more complicated and numerically unstable than standard MLT; thus,  it becomes a problematic issue when introducing this theory into stellar evolution codes. 

Several works have extended MLT to deal with convection in the presence of background chemical gradients in the past. For example, \cite{1988Ap&SS.150..115U} extended the MLT to study the impact of positive (stabilizing) background chemical gradients in the development of convective cores in massive stars. More recently,   \cite{2024ApJ...969...10C} adopted a similar model to explore convective and nonconvective motions above the crystallization front in white dwarfs. In particular, their work focuses on the final steady state that develops for given fluxes of both heat and chemical elements and they derived chemical gradients for their solutions for given chemical fluxes. In this work, we adopt a similar treatment to that of both works, essentially using the same mathematical prescription, but with a different scope and setting of the relevant physical conditions. As in \cite{1988Ap&SS.150..115U}, we are interested in the development of convection in the presence of background chemical gradients, yet we are mostly interested in  the destabilizing effects of negative chemical gradients that may develop in the late stages of low-mass stars and their consequences for the final chemical stratification and pulsations of pre-white dwarfs and white dwarfs. 
Consequently, we are mostly interested in the derivation of well-behaved equations that can be included, at minimal numerical cost, into stellar evolution computations. 

Asymptotic giant branch (AGB) stars are evolved objects with initial masses under $8\mbox{--}10 \ M_\odot$ \citep{Kipphenhahn2013,2005ARA&A..43..435H} that have already ended their helium core burning phase. They are evolved red giant stars, with an inert carbon-oxygen core surrounded by an inner helium burning shell and a more external hydrogen burning shell. The later phase of this two-shell burning configuration is not quiescent, but rather dominated by the development of events commonly termed thermal pulses. Thermal pulses are quasiperiodical thermal runaway events that develop in the helium burning shell and lead to the formation of an inner convective zone in the region located in between the two burning shells (usually named the intershell).
This short burst expands the outer layers, cooling the hydrogen burning shell, resulting in a diminution of the surface luminosity. This is followed by a longer period of the system slowly contracting again and going back to the initial configuration. Each pulse leaves behind a strong stratification in the carbon-oxygen profile, which could produce inversions of the chemical gradient leading to the development of thermohaline mixing or, alternatively, Rayleigh-Taylor instabilities and the development of chemically driven convection. This situation makes AGB stars an ideal astrophysical location for testing extensions of the MLT where chemical gradients play an important role.

Mixing processes on the AGB shape the internal structure of their descendants: white dwarfs (WDs) and pre-WDs \citep{2010A&ARv..18..471A,2019A&ARv..27....7C}. Among them, one particularly interesting population for the present study is the group of GW Vir variable stars, named after the prototype of the class PG 1159-035 \citep{1979wdvd.coll..377M,2022ApJ...936..187O}. These stars are hot, luminous stars, characterized by He-,C-, and O-rich atmospheres and by the occurrence of multiperiodic luminosity variations attributed to nonradial gravity ($g$) pulsation modes. GW Vir stars are of great interest for this work for two main reasons. First, GW Vir stars are known to display large numbers of periods corresponding to large numbers of normal oscillation modes. The presence of a large number of normal modes allows for the use of asteroseismological techniques to probe the internal chemical structure of these stars. Secondly, these stars departed the AGB only tens of thousands or even only thousands of years ago \citep{2024Galax..12...83M}. Consequently, many of the physical processes that alter the chemical stratification in older WDs, such as chemical diffusion and gravitational settling \citep{2010A&ARv..18..471A}, did not have enough time to act in these stars. This combination of a recent departure from the AGB and the large number of pulsation modes observed in their lightcurves, makes GW Vir stars excellent candidates to test the impact of convective and nonconvective mixing processes on the AGB.
Asteroseismology provides an ideal tool to unveil the internal chemical structure of stars, by comparing the observed periods with the periods predicted by theoretical models \citep{2019A&ARv..27....7C}. Moreover, the total number of observations, quantity of detected modes, and quality of  observations in GW Vir pulsators has strongly increased with the advent of space-based observatories \citep{2021A&A...645A.117C,2022ApJ...936..187O,2024A&A...686A.140C,2024A&A...691A.194C}

This work is organized as follows. In Section \ref{sec2}, we describe the MLT extension we developed to include chemically driven convection and thermohaline mixing. In Section \ref{sec3}, we describe how we included this new set of equations into a stellar evolution code to compute the evolution of AGB and post-AGB stars. We apply this prescription, compare it with the standard MLT approach and with GNA theory, and analyze the differences that arise in the chemical structure of the same star under the different prescriptions for convective (and nonconvective) motions. In Section \ref{sec4}, we explore the asteroseismological and observable consequences of these different chemical evolutions. Finally, in Section \ref{sec5}, we summarize our conclusions and discuss the future outook.

\section{A simple extension to mixing length theory} \label{sec2}

Currently, MLT is the most common and used model of convection in stellar astrophysics given to its simplicity and its good agreement with respect to observational data. Conceptually, it proposes the existence of macroscopic mass elements (or ``blobs'' of matter)
that move and transfer heat to their surroundings.
Mass elements are thought to travel a given distance (the so-called mixing length) before dissolving to the surroundings.

Convection theories are commonly used in tandem with instability criteria, which determine in the regions where turbulence develops and transports heat and/or chemical species. Radial displacements are stable against the temperature stratification \citep{Kipphenhahn2013}
  if\footnote{Note:\ in Eq. \ref{eq:stability}, the so-called
Ledoux term $B=(\varphi/\delta)\nabla_\mu$ is correct only under the assumption
of an ideal gas without strong degeneracy and coulombic interactions.} \begin{equation}
    \nabla < \nabla_\text{e} + \frac{\varphi}{\delta} \nabla_\mu ,
\label{eq:stability}    
\end{equation}
where
\begin{equation}
    \nabla\equiv \bigg(\frac{d\ln T}{d\ln P}\bigg)_\text{s},   \nabla_\text{e}\equiv \bigg(\frac{d\ln T}{d\ln P}\bigg)_\text{e}, \nabla_\mu \equiv \bigg(\frac{d\ln \mu}{d\ln P}\bigg)_\text{s},
\end{equation}
with the subindex ``e'' referring to the displaced element of stellar matter and ``s'' referring to the surroundings, respectively, and
\begin{equation}
    \varphi \equiv \bigg (\frac{\partial\ln \rho}{\partial \ln \mu}\bigg)\bigg|_{T,P} , \ \ \delta\equiv -\bigg(\frac{\partial\ln\rho}{\partial\ln T}\bigg)\bigg|_{P,\mu},
\end{equation}
where $T$ is the temperature, $P$ the pressure, and $\mu$ the mean molecular weight. It is noteworthy that even when Eq. (\ref{eq:stability}) is fulfilled, turbulence might arise. A clear example of this was demonstrated by \cite{1966PASJ...18..374K} who shown that, under the right conditions, heat diffusion can lead to the development of an oscillatory behavior with an increasing amplitude (i.e., semiconvection).

Usually, the values of the actual temperature gradient of the stellar material ($\nabla$) and convective elements ($\nabla_e$) are not known beforehand and Eq. (\ref{eq:stability}) cannot be used as a stability criterion\footnote{A notable exception is the GNA theory \citep{1993ApJ...407..284G,1996MNRAS.283.1165G} where instability regimes are determined simultaneously with the temperature gradients ($\nabla$,$\nabla_e$) and velocities ($v$).}.
Consequently, it is a standard practice to determine convective regions by checking which regions of the star cannot be stable and transport heat solely by radiation and conduction.
Moreover, since $\nabla_e$ is also not known beforehand, the most common practice is to check in which regions mass elements displaced adiabatically become dynamically unstable ($\nabla_e=\nabla_{\rm ad}$).
Under these assumptions, we can obtain the so-called Ledoux stability criterion \citep{1947ApJ...105..305L,1959PThPh..22..830S} via
\begin{equation}
    \nabla_\text{rad}<\nabla_\text{ad} + \frac{\varphi}{\delta} \nabla_\mu .
    \label{eq:false_ledoux}
\end{equation}
The radiative gradient $\nabla_\text{rad}$ is defined as the temperature gradient that would be required if the heat were to be transported only by radiation (or conduction). 
Moreover, if we neglect the impact of the chemical gradients (as the standard application of MLT does after), we then have the Schwarzschild stability criterion \citep{1906WisGo.195...41S}, namely,
\begin{equation}
    \nabla_\text{rad}<\nabla_\text{ad} .
\end{equation}
In MLT, the velocity of the convective blob \citep[][see also Appendix \ref{AppendixA}]{Kipphenhahn2013} is calculated as \begin{equation}
    v^2= g \delta (\nabla-\nabla_\text{e}) \frac{l_m^2}{8H_P} ,
\end{equation}
with $g$ being the gravitational acceleration, $l_m$ represents the ML of MLT and $H_P$ is the scale height of pressure, defined as

\begin{equation}
    H_P\equiv - \frac{dr}{d \ln P} .
\end{equation}

In MLT, we aim to obtain the values for $\nabla$ and $\nabla_\text{e}$ and, with that objective, the following two equations can be derived:

\begin{equation} \label{MLTT1}
    \nabla_\text{e}-\nabla_\text{ad} = 2U\sqrt{\nabla-\nabla_\text{e}} ,
\end{equation}
\begin{equation} \label{MLTT2}
    (\nabla-\nabla_\text{e})^{3/2}= \frac{8 U}{9} (\nabla_\text{rad}-\nabla),
\end{equation}
with
\begin{equation}
    U\equiv \frac{3acT^3}{c_P\rho^2 \kappa l_m^2}\sqrt{\frac{8H_P}{g\delta}} .
\end{equation}
In this expression, $a$ is the radiation constant, $c$ the light velocity, $c_P$ the specific heat at constant pressure, $\rho$ the density, and $\kappa$ the opacity.
In MLT equations, when the Schwarzschild criterion is considered and the usual quantities as $\nabla_\text{rad}$ and $\nabla_\text{ad}$ are given, we can obtain unique real solutions for $\nabla$ and $\nabla_\text{e}$ and  the convective velocity, $v$, can \textbf{}be obtained on their basis. However, if we consider the Ledoux criterion instead of a Schwarzschild,  a region that is Ledoux unstable but Scharzschild stable\footnote{Also, a region Ledoux stable appears in Schwarzschild unstable dominion, and considering double-diffusive mixing is where semiconvection takes place. In this work we will not consider semiconvection motions and we will treat such regions as convectives using MLT, except when using GNA where semiconvection is included.} appears. The new unstable region corresponds to a Rayleigh-Taylor (RT) instability.

We can obtain expressions similar to Eqs. (\ref{MLTT1}) and (\ref{MLTT2}) without neglecting the impact of the chemical gradients to describe the convective motions. The modified equations for $\nabla$, $\nabla_\text{e}$, and $v$ are (see Appendix \ref{AppendixA}) as follows,

\begin{equation} \label{MLT+1}
    \big(\nabla_\text{e}-\nabla_\text{ad}\big)\big(\nabla-\nabla_\text{e}-\frac{\varphi}{\delta}\nabla_\mu\big)^{1/2}=2U\big(\nabla-\nabla_\text{e}\big) ,
\end{equation}
\begin{equation} \label{MLT+2}
    \big(\nabla-\nabla_\text{e}-\frac{\varphi}{\delta}\nabla_\mu\big)^{1/2}(\nabla-\nabla_\text{e})=\frac{8}{9}U\big(\nabla_\text{rad}-\nabla) , 
\end{equation}
\begin{equation} \label{MLT+3}
    v^2 = g\delta \bigg(\nabla - \nabla_\text{e} - \frac{\varphi}{\delta}\nabla_\mu\bigg)\frac{l_m^2}{8H_P}.
\end{equation}

This set of equations is equivalent to Eq. (25) of \cite{1988Ap&SS.150..115U} and Eq. (20) of \cite{2024ApJ...969...10C}. Neither of these works used the derived mixing velocities and temperature gradients as part of stellar evolution computations but to study specific processes. On the one hand, \cite{1988Ap&SS.150..115U} studied to which extent convection was possible in the presence of positive (stabilizing) chemical gradients. They concluded that in the presence of finite thermal diffusion (regardless how trivial),  the instability criterion derived under the assumption of adiabatic perturbations overestimates the instability region.
On the other hand,     \cite{2024ApJ...969...10C} studied the expected steady state that develops in the interior of a white dwarf, under the assumption of given values of the heat and chemical fluxes.
Our objective, however, is to use Eqs. \ref{MLT+1} and \ref{MLT+2} as part of a stellar structure and evolution code to compute the mixing and heat transport in the presence of chemical gradients caused by nuclear burning. We refer to this simple extension as MLT$\sharp$ and its derivation is described in Appendix \ref{AppendixA}. 

  \begin{figure*}
   \centering
   \includegraphics[width=\textwidth]{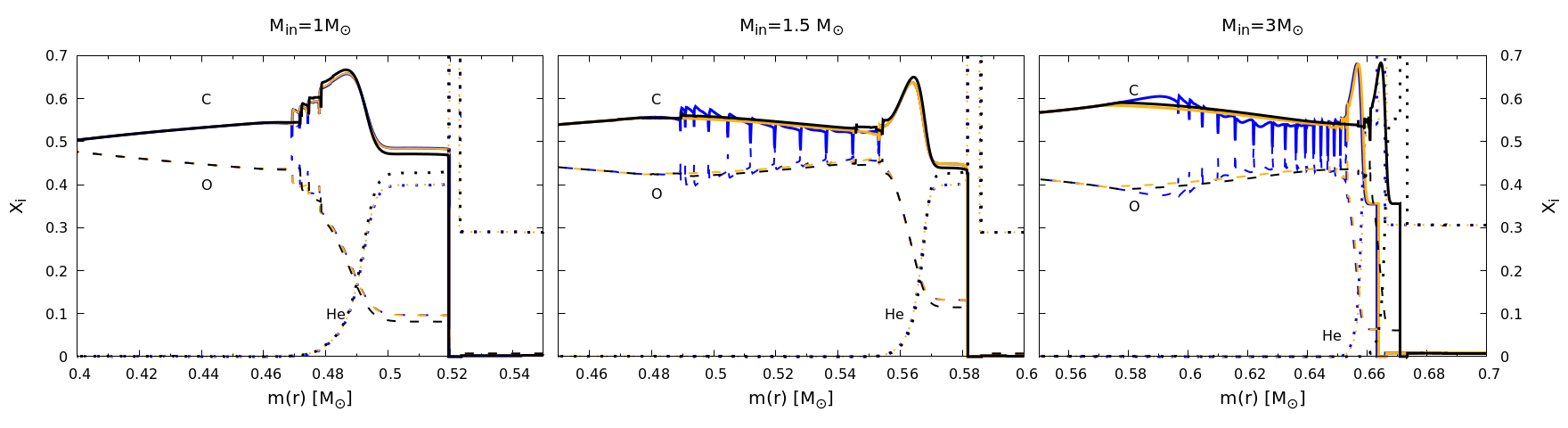}
   \caption{Chemical abundances of He (dotted lines), C (continuous lines) and O (dashed lines) of stellar models of initial mass $M_\text{in}=1M_\odot$ (left panel), $1.5M_\odot$ (middle). and $3M_\odot$ (right) at the end of the AGB phase. The blue lines correspond to the case where MLT was used, the orange ones are for  MLT$\sharp$, and the black ones  are for GNA.}
              \label{Fig1}%
    \end{figure*}

Having obtained equations for convective motions considering Ledoux criterion and the effect of chemically driven convection (RT instabilities), we still need to consider one more scenario. First, we know that a negative $\nabla_\mu$ will favor mixing due to the tendency of the heavier layer to sink under the lighter one. However, regions with $\nabla_\mu<0$ and Ledoux stable are possible when the heavier element is sufficiently hotter than the light layer below. While it is dynamically stable, over time, thermal diffusion will act, cooling the blob, and the element will then sink and mix with the surrounding environment. This is the regime of thermohaline mixing and this process  is significantly slower than dynamic convection. Nonetheless, the timescales are still much shorter than typical stellar evolution timescales and, thus, thermohaline mixing becomes relevant, for example, in stars on the red giant branch (RGB; \citealt{2011A&A...533A.139W}) and during the thermal pulses on the AGB.

It is noteworthy that in this model, Eqs. (\ref{MLT+1}) and (\ref{MLT+2}) give solutions beyond the limit established by the Ledoux criterion, entering the regime of thermohaline mixing, as long as $\nabla-\nabla_\text{e}-\frac{\varphi}{\delta}\nabla_\mu>0$ and $\nabla_\text{rad}<\nabla_\text{ad}$. However, \cite{2024ApJ...969...10C} showed that when assuming inefficient convection, the mixing velocity of the extended MLT model in the sub-Ledoux regime takes the same form as in the parametrization given by \cite{1980A&A....91..175K} for thermohaline mixing. Below the critical line defined by Eq.  \ref{eq:false_ledoux}, we find that Eqs. \ref{MLT+1} and \ref{MLT+2} might have multiple solutions, ranging from an almost fast-adiabatic convection to a slow thermohaline behavior with $\nabla\simeq\nabla_{\rm rad}$. In these circumstances we decided for the most conservative approach and decided to adopt the thermohaline solution for all cases below the critical line given by Eq. \ref{eq:false_ledoux} to deviate the least from previous works. We will employ then, for this region of the parameter space, the following expression given by \cite{1980A&A....91..175K} for the mixing velocity,
\begin{equation} \label{THERMOV}
    v_T=\frac{6acT^3}{c_P\rho^2\kappa}\frac{\frac{\varphi}{\delta}\nabla_\mu}{(\nabla-\nabla_\text{ad})}
.\end{equation}
For the temperature gradient, in this case, we consider $\nabla=\nabla_\text{rad}$.

As stated before, both RT and thermohaline instabilities are considered stables under Schwarzschild criterion (SC) and, then, standard MLT will not mix regions that are neither RT, nor thermohaline.

The extension of the standard MLT that we present in this paper, which we refer to as MLT$\sharp$ is defined by the  following prescriptions: (i)  standard MLT when the regions are Schwarzschild unstable; (ii) Eqs. (\ref{MLT+1}), (\ref{MLT+2}), and (\ref{MLT+3}) when the region is Schwarzschild-stable, but Ledoux-unstable (RT instabilities); and (iii) thermohaline mixing given by Eq. (\ref{THERMOV}) when the layer is Ledoux-stable, but $\nabla_\mu<0$. In the next section, we describe how we applied the MLT$\sharp$ model in AGB stars and we compare the results with using standard MLT and GNA prescriptions.

%


\section{Mixing processes in the asymptotic giant branch} \label{sec3}

\begin{figure*}
    \centering
    \includegraphics[scale=0.4]{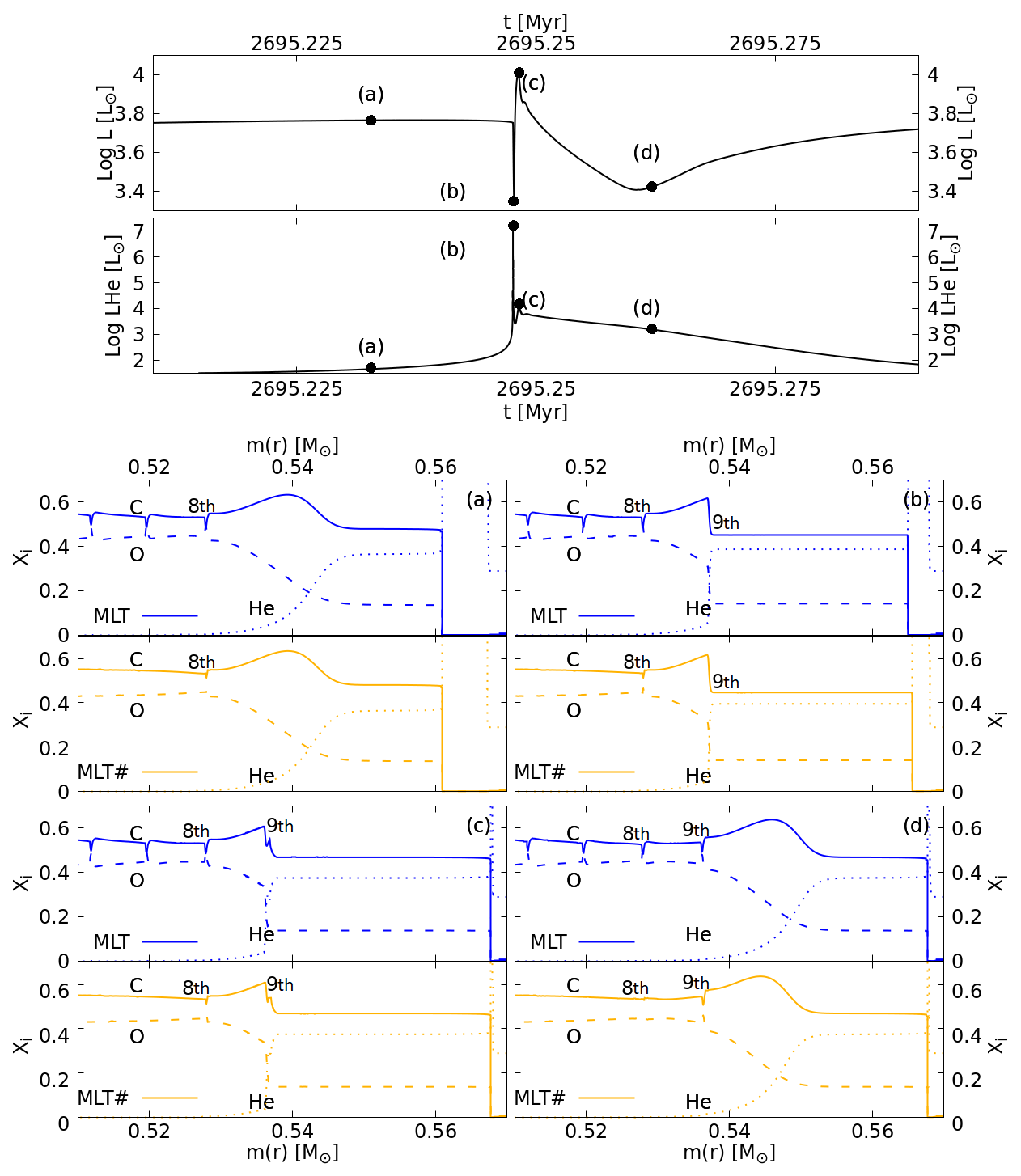}
    \caption{\textit{Top two panels}: Surface luminosity (top) and helium luminosity (bottom) during the ninth thermal pulse of the sequence with $M_\text{in}=1.5 \ M_\odot$. Four stages, labeled as \textit{a, b, c,} and \textit{d}, are highlighted, which correspond to the chemical stratifications shown in the four bottom panels. \textit{Bottom panels:} Formation of the O peak during the thermal pulses. Panels \textit{b} and \textit{c} show that the O peak is formed in the same way using both prescriptions, while panel \textit{d} shows how the additional instabilities considered in MLT$\sharp$ lead to the mixing of the O peak formed in the previous TP. We also highlight the position of the eighth and ninth O peaks.}
    \label{figpulso}
\end{figure*}

After the helium core-burning phase, stars with initial masses of $M_\text{in}<8  \text{--} 10M_\odot$ move to the AGB in the Hertzprung-Russell (HR) diagram \citep{Kipphenhahn2013}. These stars posses an inert carbon(C)-oxygen(O) core surrounded by two concentric shells where helium and hydrogen burn. 
The advancing He-burning shell becomes progressively thinner, leading to an unstable condition. This unstable setup will lead to the development of thermal instabilities in the He-burning shell.
These thermal instabilities, known as thermal pulses (TP), lead to the formation of inner convective regions where the material is periodically mixed.
Each thermal pulse will leave behind a chemical stratification with a sharp O abundance peak (Fig. \ref{Fig1}, blue lines) at the outer regions of the CO core \citep{2005ARA&A..43..435H,Kipphenhahn2013}.

These peaks in the O abundance could lead to inversions in the chemical gradient and then those layers in the star will be either RT unstable or unstable against thermohaline processes. As stated before, these regions are not mixed in standard stellar evolution computations (MLT+SC). This scenario is then ideal for testing the development of RT or thermohaline instabilities and to study the impact of chemically driven convection (CDC) and thermohaline mixing in evolved stars. Moreover, the existence of variable stars immediately after this stage (the GW Vir stars mentioned in the introduction) opens up the possibility to test these predictions and, in the future, calibrate the intensity of those mixing processes.

To explore the impact of mixing processes  induced by negative chemical gradients, we carried out simulations for three different initial masses: $M_\text{in}=1 \ M_\odot, \ 1.5 \ M_\odot, \ \text{and} \ 3 \ M_\odot$ using \texttt{LPCODE} \citep{2016A&A...588A..25M,2020A&A...633A..20A}. Since we are interested only in the physical processes that take place during the AGB phase\footnote{As stated before, thermohaline and RT instabilities can appear in earlier stages like the RGB phase, but we are not going to study such processes in this work.}, all the prior evolution was carried using standard MLT and Schwarzschild criterion to avoid numerical discrepancies. When the stars were located into the AGB, we split the sequences into three different paths: one with standard MLT, one with MLT$\sharp$, and a third with GNA. The GNA test was done  with the purpose of comparing the outcome of the simple MLT$\sharp$ with that of a significantly more complex (and, thus, more complicated and numerically problematic) theory.
Moreover, in this paper, we limit our analysis to the outer regions of the core and the intershell. Possible mixing processes at the innermost regions of the CO core immediately after the end of core He burning will be addressed in future works. We note here that both GNA and the current version of MLT$\sharp$ (which considers $\frac{\varphi}{\delta}\nabla_\mu$ and not the complete Ledoux term $B$) are inadequate when strong degeneracy of the electron gas and nonideal interactions become important.

In all the sequences, we considered an initial metallicity of $Z=0.015$ and mixing length of $l_m=1.822$. The rest of our choices for the input physics are the same as in \cite{2016A&A...588A..25M}.

As stated before, during each thermal pulse, a discontinuity in the carbon-oxygen chemical profile is created, which is peak-shaped as we can see in Fig. \ref{Fig1} (see the blue lines). However, this does not directly imply an inversion of the chemical gradient in the region, as long as a remnant abundance of helium capable of stabilizing those layers is present. Only after the helium in this region is burned $\nabla_\mu$ becomes negative and MLT$\sharp$ and GNA mix the elements and smooth the profiles. MLT never mix those regions, as explained before. In Fig. \ref{figpulso} we show a comparison of the chemical evolution using MLT and MLT$\sharp$ during the same TP for the case of $M_\text{in}= 1.5 \ M_\odot$. We find that the timescale for the mixing of an O-peak in the simulations is roughly similar to the duration of a single TP cycle. From Fig. \ref{figpulso}, we can see that the mixing in the ninth TP for the sequence with $M_\text{in}=1.5 \ M_\odot$ occurs in less than 50000 years.  

Regarding the effects of the MLT$\sharp$ and GNA prescriptions in the cases studied here ($M_\text{in}= 1 \ M_\odot, \ 1.5 \ M_\odot, \ \text{and} \ 3 \ M_\odot$), we can see some differences between them. In the $M_\text{in}=1M_\odot$ sequence, the star was characterized by four TP before leaving the AGB. 
This value did not allow for a complete burning of He in the region where the O-peaks were located and, consequently, these regions remained stable. Had the star prolonged its AGB  phase
(e.g., with a slower rate of mass loss), the remaining He would have been burned and the mixing induced by the negative chemical gradients would have started. The sequences with $M_\text{in}=1.5M_\odot$ and $M_\text{in}=3M_\odot$ were characterized by 11 and 16 TP, respectively, more than enough to burn the He tail and mix the carbon-oxygen profiles. Furthermore, as the temperature of the helium remnant region rises with the mass of the star, they need fewer TP to start the process. In the present computations, the $M_\text{in}=1.5M_\odot$ sequence started the mixing in the fourth TP and the $M_\text{in}=3M_\odot$ sequence burned the remaining helium tail  as early as in the second TP.
As the burning shell advances, the remaining He tail is burned and mixing erases the O-rich peaks. At any given time, this means that only the O-rich peaks created during the last TP (or last two TPs) will still be present in the interior of the model  (see middle panel in Fig. \ref{Fig1}). Consequently, after the star departs from the TP-AGB only the later TP-induced O peaks will still be present in the model.

\begin{figure}
    \centering
    \includegraphics[scale=0.35]{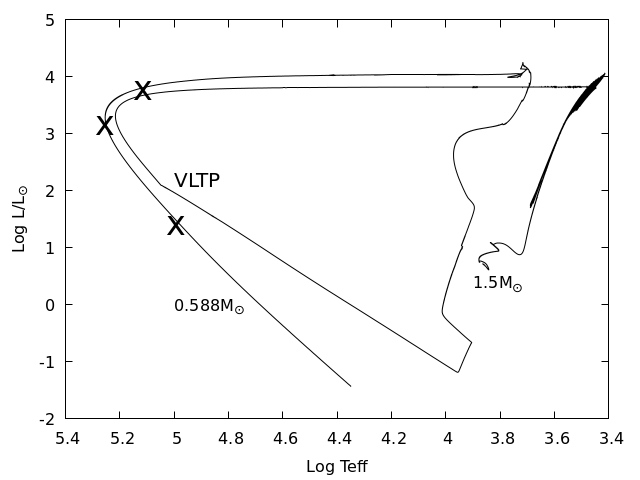}
    \caption{Evolutionary track for the $M_\text{in}=1.5M_\odot$ star. The VLTP is highlighted as well as the three points (marked with crosses), where the pulsational calculations were made.}
    \label{FigHR}
\end{figure}

 \begin{figure*}
   \centering
   \includegraphics[width=\textwidth]{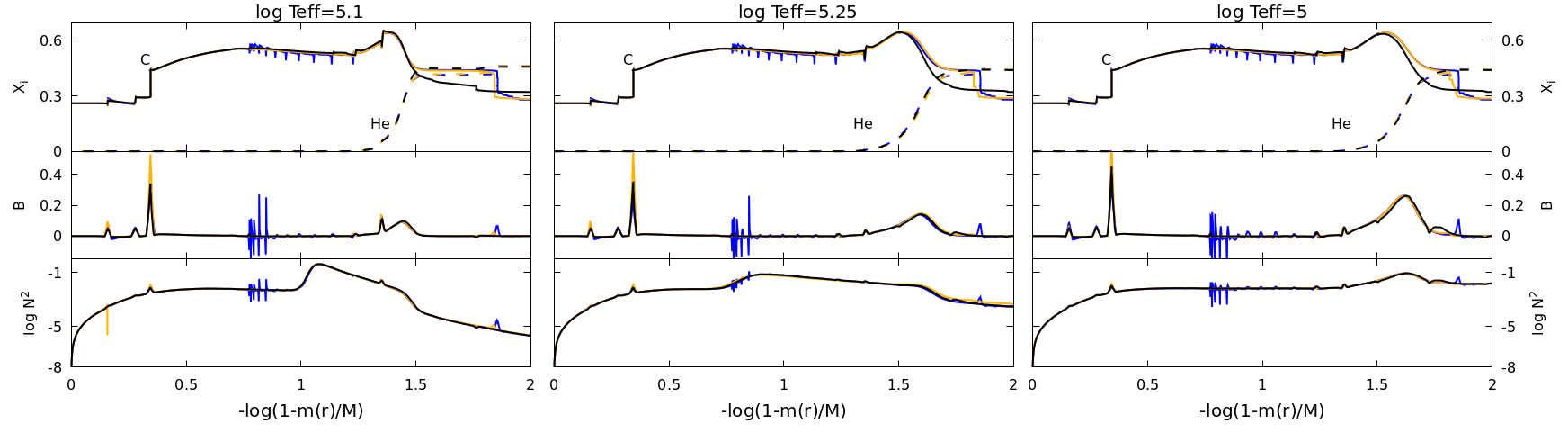}
   \caption{Chemical profiles (top), Ledoux term B (middle), and the logarithm of the squared Brünt-Väisälä frequency (bottom) for a PG1159 model with a mass of $M=0.58 \ M_\odot$ ($M_\text{in}=1.5 \ M_\odot$) at three different stages after the VLTP.  As in Fig. \ref{Fig1}, blue lines correspond to MLT, orange lines to MLT$\sharp$, and black lines to GNA.}
              \label{FigBN}%
    \end{figure*}

We can make one last observation regarding the $M_\text{in}=3M_\odot$ sequence. While in the other two sequences the three chemical profiles have the same overall shape (except for the peaks in MLT) and they have, for example, the same place and height for the maximum in the carbon abundance; however, this is not the case for the more massive star. The GNA profile appears to be shifted towards the exterior.
This shift is a direct consequence of the stabilizing effect of the positive chemical gradient at the bottom of the H-rich envelope that weakens the intensity of the third dredge up episodes. Such a stabilizing effect is only taken into account in GNA simulations, where the instability criterion is self-consistently based on Eq. (\ref{eq:stability}). We note that our simulations do not consider any kind of convective boundary mixing at the bottom of the convective envelope. Such nonlocal mixing process is known to favor third dredge up and would erode the chemical transition and remove any stabilizing effect of $\nabla_\mu$ at the bottom of the H-rich envelope \citep[e.g.,][]{2000A&A...360..952H,2006NuPhA.777..311S,2015ASPC..493...83M, 2016A&A...588A..25M, 2020MNRAS.493.4748W}. In our setup with $M_\text{in}=3M_\odot$, this
resulted in the third dredge-up starting one TP earlier and also in stronger dredge-up episodes in MLT and MLT$\sharp$ sequences with respect to GNA ones.

\section{Pulsational properties of PG1159 stellar models} \label{sec4}

In the previous section, we showed that the choice of different prescriptions for the treatment of stellar convection can lead to significant differences in the final chemical profile of a stellar model. As the intensity of those mixing processes are badly understood, we might wonder whether the predictions from GNA and MLT$\sharp$ prescriptions can be tested in real stars. One way to test these chemical structures is by studying GW Vir stars with asteroseismological techniques \citep[see][and references therein]{2019A&ARv..27....7C}.

As mentioned in the introduction, GW Vir are pulsating pre-WDs and WDs with PG1159 spectral types \citep{2006PASP..118..183W}. The vast majority of PG1159 stars are thought to have formed after a very late thermal pulse (VLTP)\footnote{ Noting a minority of low-luminosity member might be the result of very peculiar mergers \citep{2022MNRAS.511L..66W,2022MNRAS.511L..60M}}. This process takes place once the star has departed from the AGB and is entering the WD cooling stage \citep[see][for a review on different evolutionary channels]{2024Galax..12...83M}. To produce appropriate stellar models of PG1159 stars, compute their pulsational properties and test the impact of different chemical profiles, we need to continue the evolution of the sequences computed in the previous section through their departure from the AGB and through a VLTP. In this first exploratory study, we have focused on the three $M_\text{in}=1.5M_\odot$ sequences computed with the three different convection theories. We have tuned the mass loss at the end of the AGB in a way so that the model departs at the right time for a VLTP to occur. We then followed the evolution through the He-flash and the consequent proton ingestion event. Figure \ref{FigHR} shows the resulting evolution on the HR diagram. Following the VLTP, the star evolves back to the giant region of the HR diagram and it later contracts again to the WD cooling track, now as a H-deficient PG1159 star. At this stage, we took three different snapshots of the stellar structure and computed their adiabatic $g$-mode periods (crosses in Fig. \ref{FigHR}). Those snapshots correspond to $\log T_\text{eff}=5.1$, before the knee in the HR diagram, the point of maximum effective temperature at $\log T_\text{eff}=5.25$ and, finally, $\log T_\text{eff}=5$ after the knee, at the entrance of the WD cooling phase.

Adiabatic pulsations of our PG1159 models were computed using an updated version of  \texttt{LP-PUL} \citep{2006A&A...454..863C}. Among other quantities, \texttt{LP-PUL} computes the pulsation periods  $\Pi_{k,\ell}=2\pi/\sigma_{k, \ell}$, with $k$ being the radial order and $\ell$ the harmonic degree of a given normal mode. In this work, we limited our analysis to dipolar $g$-modes ($\ell=1$). To get the values of periods as precise as possible, we have employed about 5000 mesh-points to describe background stellar models. In pulsating stars, in general, the structure of the $g$-mode period spectrum is extremely sensitive to the radial profile of the squared value of the Brunt-Väisälä (or buoyancy) frequency ($N$). In order to accurately compute the value of $N$ in the core of PG1159 stars, we used the treatment described by \cite{1990ApJS...72..335T, 1991ApJ...367..601B}, where the Brunt-Väisälä frequency is given by
\begin{equation}
    N^2 = \frac{g^2 \rho}{P} \frac{\chi_T}{\chi_\rho} \big ( \nabla_\text{ad} - \nabla - B \big ).
    \label{eq:BV2}
\end{equation}
In this expression, $B$ is the so-called Ledoux term \citep{1990ApJS...72..335T}, expressed as
\begin{equation}
    B= - \frac{1}{\chi_T} \sum_1^{M-1}\chi_{X_i} \frac{d\ln X_i}{d \ln P},
\end{equation}
while the terms $\chi_\rho$, $\chi_T$, and $\chi_{X_i}$ are
\begin{equation}
    \chi_T= \bigg (\frac{d\ln P}{d \ln T}\bigg )_{\rho,\{X_i\}} , \ \chi_{X_i}=\bigg(\frac{d\ln P}{d \ln X_i}\bigg)_{\rho,T,\{X_{j\neq i}\}},\ \
    \chi_\rho=\bigg (\frac{d\ln P}{d \ln T
    \rho}\bigg )_{T,\{X_i\}}.
\end{equation}

\begin{figure}
   \includegraphics[width=\columnwidth]{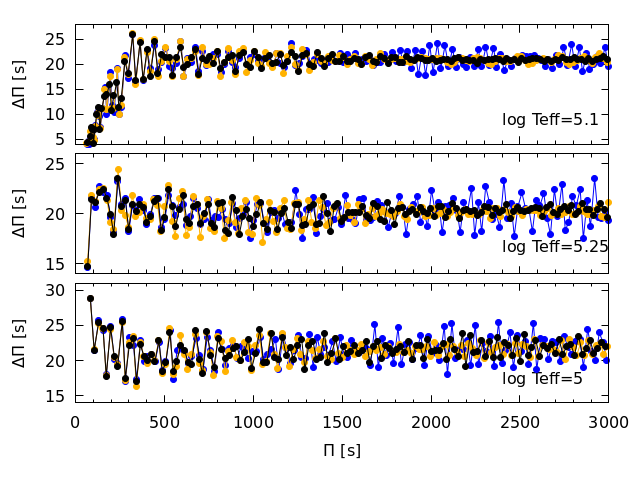}
      \caption{Forward period spacing for the PG1159 models with masses of $M=0.58 \ M_\odot \ (M_\text{in}=1.5 \ M_\odot )$ and the three mixing prescriptions at the three different stages selected after the VLTP. As in the previous figures, blue dots correspond to MLT, orange dots to MLT$\sharp$, and black dots to GNA.}
         \label{FigDP}
\end{figure}

Figure \ref{FigBN} shows the internal chemical stratification of our computed PG1159 models for the main nuclear species (upper panels), together with the values of the Ledoux term (middle panels), and the squared Brunt-Väisälä frequency (lower panels). The figure emphasizes the role of the chemical interfaces on the shape of the Brunt-Väisälä frequency profile. In fact, each chemical transition leads to distinctive features in $N^2$, which are eventually responsible for the mode-trapping properties of the model \citep{2019A&ARv..27....7C}. In the inner part of the core ($q=-\log 1-m/M_\star < 0.5$), there are several peaks in $B$ (and $N^2$) that are related to changes in the mass fractions of C and O created immediately after the end of the He-core burning stage.  In the outermost part of the CO core ($0.7<q<1.5$), we can see the abrupt features shaped during the TP-AGB phase. They are specially pronounced in the case of the model computed under the standard assumptions (MLT+SC, blue lines in Fig. \ref{FigBN}). In this case, we can see how the O-peaks created during the thermal pulses lead to the formation of clear spikes in $N^2$. These spikes are not present in those models computed with either the GNA or MLT$\sharp$ approaches, where these regions are mixed by either RT instabilities or thermohaline processes. We expect these spikes to lead to clear mode-trapping features of $g$-modes.

\subsection{Mode trapping}

Mode trapping is a phenomenon that occurs when a star has abrupt structural features (e.g., sharp chemical transitions) that cause the amplitudes of the pulsation modes to be significantly different from those that would occur if such features were not present. As a consequence, the period differences between consecutive periods are not constant;
rather, they vary significantly, having pronounced minima which recur as the period
increases (i.e., the “trapping cycle”), as described by \cite{2005ASPC..334..553M,2019A&ARv..27....7C}. In WD stars, the structure of the $g$-mode period spectrum is sensitive to the spatial profile of the Brunt-Väisälä frequency \citep{2019A&ARv..27....7C}. A clear signature of mode trapping is that the forward period spacing (defined as $\Delta \Pi_k=\Pi_{k+1}-\Pi_k$, as a function of the pulsation period, $\Pi_k$) displays strong deviations from a constant value.
Moreover, this quantity has been shown to be sensitive to the chemical composition of the star, where chemical gradients in different regions of the star affect the period spacing at different ranges of periods \citep{2006A&A...454..863C}.
In light of the clear differences in the squared Brunt-Väisälä frequency of the different models (shown in Fig. \ref{FigBN}), we would expect models computed with the standard MLT to display larger mode trapping features than those computed under the assumption of either GNA or MLT$\sharp$. In Fig. \ref{FigDP}, we show $\Delta \Pi_k$ versus $\Pi_k$  for PG1159 models where the MLT, MLT$\sharp$, and GNA prescriptions have been employed at the three different stages given in Figs. \ref{FigHR} and \ref{FigBN}.
In Fig. \ref{FigDP}, we shows that for periods longer than $\sim 1700$ s,  MLT models (in blue) have much larger deviations from the mean period spacing than both GNA and MLT$\sharp$ models. This is true for the three evolutionary stages depicted in the figure. 

 To ensure that the broad deviations in the constant period spacing observed in the MLT models are indeed coming from the O-peaks created by the thermal pulses (at $0.7<q<1.5$, Fig. \ref{FigBN}), we performed a similar analysis to the one described in Section 3.4 of \cite{2006A&A...454..863C}. This consisted of artificially suppressing the Ledoux term $B$  in different parts of the star. One of the advantages of using Eq. (\ref{eq:BV2}) is that the contribution to $N^2$ from changes in the chemical composition is almost completely contained in the Ledoux term $B$. This fact renders the method particularly useful for inferring the relative weight that each chemical transition region has in the mode-trapping properties of the models. With this in mind, we divided the stellar models into three regions, corresponding to the core ($q<0.7$), O-peak region ($0.7<q<1.45$), and the envelope ($q>1.45$).

\begin{figure}

   \includegraphics[width=\columnwidth]{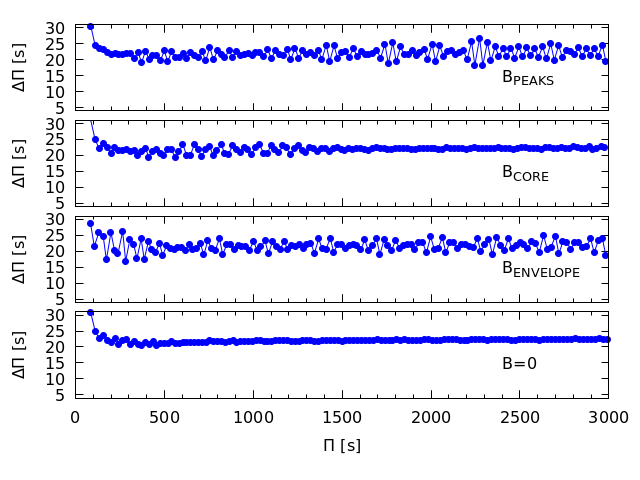}
      \caption{Forward period spacing versus periods for the MLT case and $\log T_\text{eff}=5$ when the Ledoux $B$ term has been artificially suppressed in different regions and  does not remain null in the regions specified by the different subscripts of $B$.  The only exception is where  $B$ has been suppressed for the whole model  (\textit{bottom}), labeled as $B=0$. We see  that, when $B$ is only considered not-null in the regions of  peaks of the C-O profile, several periods are affected, with the most important contributions being for the periods from $1700$ s and higher (\textit{top}).}
         \label{FigDPS}
   \end{figure}

In Fig.  \ref{FigDPS}, we show the resulting period spacing for the model with $\log T_\text{eff}=5$ and the MLT prescription, when $B$ is set to $B=0$ outside each predefined region. Thus, in the first panel we show the case when the Ledoux term is only considered for the region of the peaks, the second panel considers only the Ledoux term in the core, and the third only in the envelope. Finally, for comparison, the fourth panel sets $B=0$ in the whole stellar model, showing that the deviations from a constant forward period spacing seen in the previous panels and in Fig. \ref{FigDP} are consequences of the non-null Ledoux term. As we can see, Fig.  \ref{FigDPS} shows clearly that the core chemical transitions mostly affect mode trapping for $\Pi <1500$s while the O-peaks, created by the thermal pulses, lead to a distinctive trapping feature at longer periods. Envelope chemical transitions lead to mode trapping features over the whole range of plotted periods. In light of the similar chemical structure of the envelopes of MLT, MLT$\sharp$ and GNA models, Fig. \ref{FigDPS} shows that the much larger trapping features observed in Fig. \ref{FigDP} for $\Pi > 1700$s are indeed caused by the O-peaks left behind by the standard MLT treatment. This could be thought of as a useful seismic tool for deducing whether chemically induced mixing processes are present in the interiors of actual stars.

\begin{figure}

   \includegraphics[width=\columnwidth]{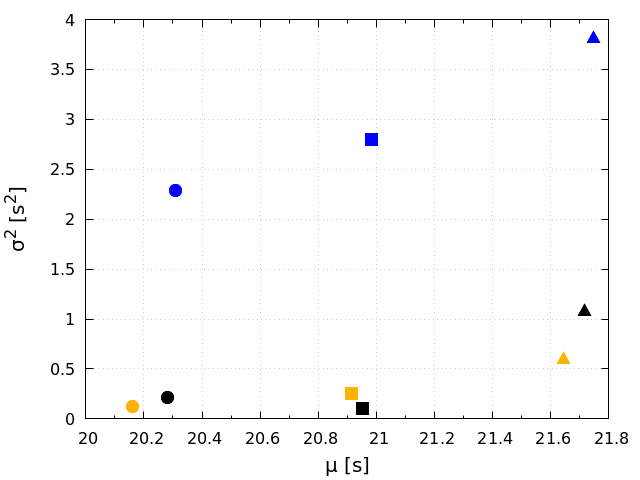}
      \caption{Mean value $\mu$ and variance $\sigma^2$ of the forward period spacing using periods $\Pi>1700$ s. The squares represent the stage of $\log T_\text{eff}=5.1$, the circles are $\log T_\text{eff}=5.25$, and the triangles are $\log T_\text{eff}=5$. MLT shows significantly greater variances in all three stages}         \label{FigVar}
   \end{figure}

Finally, having identified the most significant contributions for the peaks in the period spacing, we can go back to Fig. \ref{FigDP} and test the changes in the amplitudes of $\Delta\Pi$ statistically, for $\Pi>1700$ s. Thus, we carried out an estimation for the mean value and the variance of the $\Delta \Pi$, which are shown in in Fig. \ref{FigVar}. We obtained much higher variances for MLT and similar variances between MLT$\sharp$ and GNA.

\subsection{Comparison with GW Vir stars}

\begin{figure*}
    \centering
    \includegraphics[width=\textwidth]{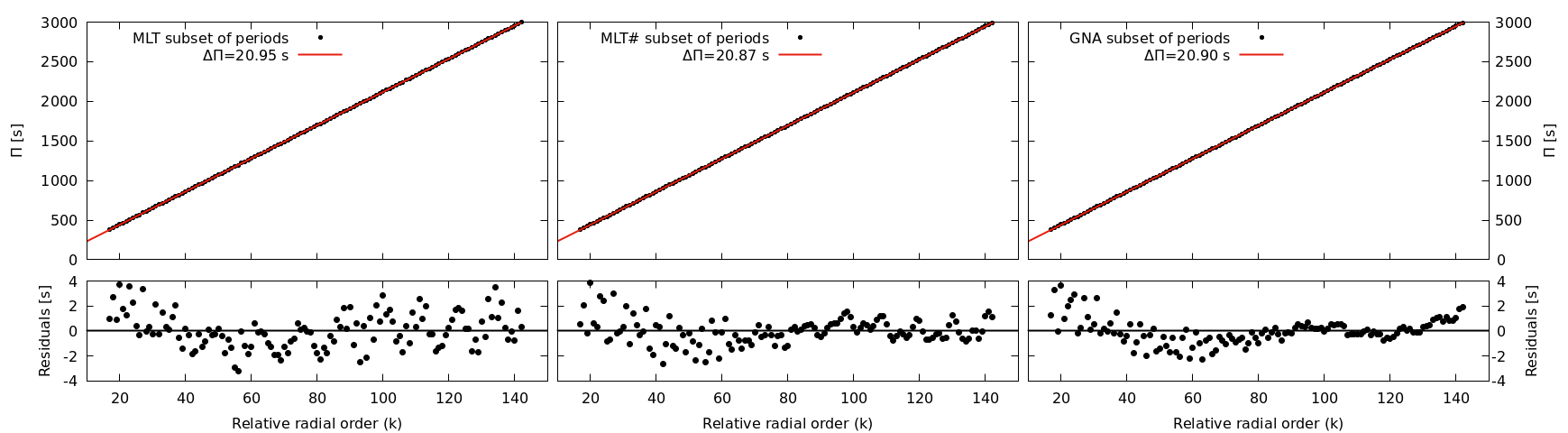}
    \caption{Linear period fit (upper panel) and residuals (lower panel) for the $\log T_\text{eff}=5.1$ models computed under the three different convective theories.}
    \label{figres}
\end{figure*}

\begin{figure}
   \includegraphics[width=\columnwidth]{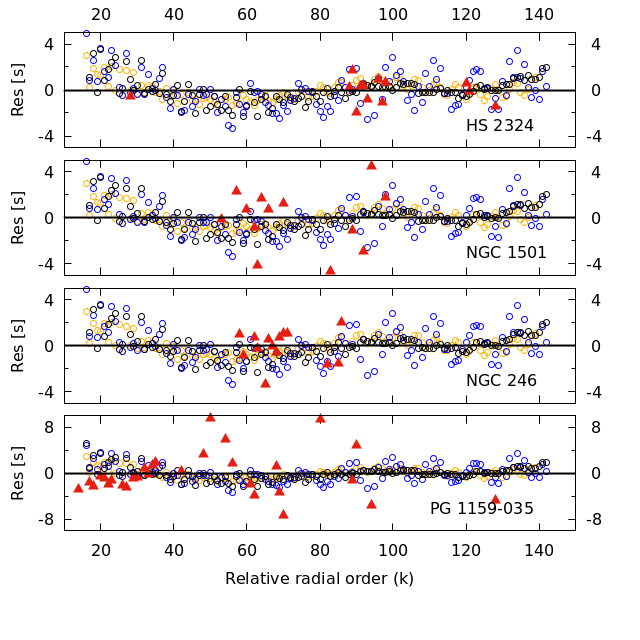}
    \caption{Residuals from the theoretical models (empty circles) and the observed pulsating stars (red triangles). As in the previous figures, blue corresponds to MLT, orange to MLT$\sharp,$ and black to GNA. There is a notable change in scale with respect to the comparison with PG 1159-035 \textit{(bottom)} due to its larger residuals.}
    \label{figresiduals}
\end{figure}

The computation of the forward period spacing in actual stars requires the observation of normal modes of consecutive radial order, which is not always possible. Therefore, when comparing with real pulsating stars, it is often more useful to rely on the residuals of the periods. Residuals are defined as the difference between the observed periods, $\Pi_i^\text{O}$, and the periods derived from a mean period spacing, $\Pi^\text{fit}_i$ \citep{2007A&A...475..619C}. The mean period spacing is obtained making a linear least squares fit to the period sample. 
We know  from both theory and observations \citep{2006A&A...458..259C, 2021A&A...645A.117C} that only the high luminosity GW Vir stars (i.e.m before the "knee" in the HR diagram, see Fig. \ref{FigHR}) show pulsation periods longer than 1000 s. Therefore, given that the O-peaks created during the thermal pulses only affect the longer periods, for this exploratory work we will limited our comparisons to the model with $\log T_\text{eff}=5.1$ (see Fig. \ref{figres}). We can see that the main differences between the residuals in MLT and MLT$\sharp$ and GNA arise for higher orders of $k$, having MLT significantly higher residuals in that regime.

Finally, we aim to test our models with actual observations of pulsating stars. The relevant period range for our work corresponds to periods longer than $1700$ s, therefore, we  focused our comparisons with GW Vir stars that exhibit most of their periods in that particular range. We selected the following stars: PG\,1159-035 \citep{2022ApJ...936..187O}, NGC\,1501, HS\,2324 \citep{2021A&A...645A.117C}, and NGC\,246 \citep{2024A&A...686A.140C}. We replicated the average period spacing found in the references and then proceeded to estimate and compare the residuals with the ones from our simulations. We show the comparison in Fig. \ref{figresiduals}. Given theoretical models are not asteroseismic models, the comparison
is only qualitative; in other words, they are not the result of period-to-period fits to the observations, but instead they represent snapshots of theoretical models in a similar evolutionary stage to that of the observed stars.

The most notable feature in Fig. \ref{figresiduals} is the extremely large residuals found in PG\,1159-035 (i.e., up to 10 s). Such a large extent of mode trapping is far greater than anything displayed by our models in any period range. Given that chemical transitions in the envelope would be able to produce such huge trapping features, we suspect that these high residuals values are connected to a different envelope structure than those corresponding to our models. This would be in line with the suggestion by \cite{2008ApJ...677L..35A}, based on the measured rates of period changes, that PG\,1159-035 harbors an envelope much thinner (in terms of mass) than predicted by standard stellar models.
For the other stars, the comparison is inconclusive. On the regime of interest, $k> 80$, HS 2324 shows, on average, residuals that are in line with those observed in our MLT$\sharp$ model, notably smaller than those of our standard MLT computations. The residuals of NGC\,1501 and NGC\,246 are more in line with the predictions of the MLT models, yet the number of periods in the long period regime is rather small. Further studies, based on proper asteroseismic models of the relevant stars, are needed to reach any conclusion that would favor  any of the mixing treatments considered in this work and to confirm, for example, whether thermohaline mixing processes or RT instabilities can take place during the TP-AGB phase.

\section{Conclusions} \label{sec5}

In this work, we derived an extension to standard MLT that incorporates the existence of chemical gradients in the background stratification, which we named MLT$\sharp$ (see Appendix \ref{AppendixA}). In our formulation, we included the prediction of the double diffusive convective theory GNA, developed by \cite{1996MNRAS.279..305G}. Next, we explored the importance of chemically driven convection and thermohaline processes during the TP-AGB phase and explored their consequences for pulsating GW Vir stars.

Using \texttt{LPCODE}, we computed the evolution of stars with initial masses $M_i=$1, 1.5 and 3 $M_\odot$ during the TP-AGB phase where thermal pulses create layers of enhanced O-rich composition (O-peaks) and inversions of the chemical gradients may develop. For each initial mass, we computed three different TP-AGB evolutionary sequences, with the standard MLT, with MLT$\sharp$, and with the GNA. 
We find that when enough thermal pulses take place on the TP-AGB, negative chemical gradients establish that lead to the development of both chemically driven convection and thermohaline processes. In our simulations, this happens for the more massive sequences in our sample ($M_i=$ 1.5 and 3 $M_\odot$). Both MLT$\sharp$ and GNA predict that the mixing timescales for these instabilities are shorter than the length of the TP-AGB phase itself.
Consequently, when these processes are taken into account, the final chemical stratification in the outer regions of the CO cores of these models are smoother than those arising from stellar evolution computations that adopt the standard MLT prescription. 
On the contrary, our $M_i=1 M_\odot$ sequences never develop thermohaline or Rayleigh-Taylor instabilities; in addition, all sequences retain a chemical profile with several O-peaks in the outer core, regardless of the adopted convective theory.

We also explored the possibility of using pulsating H-deficient pre-WD stars (GW Vir) to establish the prevalence of these O-peaks in real stars. In this connection, we computed the evolution of our three $M_i=$ 1.5$M_\odot$ sequences through a final post-AGB thermal pulse to construct models corresponding to GW Vir stars. Once they were constructed, we computed their adiabatic pulsational spectrum and studied their period spacing features. 

We find that the O-peaks created during the TP-AGB produce mode-trapping features in the period spacing for modes with $\Pi>1700$ s. Moreover, in such period range, the mode trapping features caused by the chemical transitions in the inner CO core are irrelevant. We do find, however, that chemical transitions in the envelope can lead to mode trapping features in the whole range of studied periods. Our models were computed with the standard MLT approach and, therefore,  they include strong O-peaks, leading to much larger trapping features at $\Pi>1700$ s than those computed with either MLT$\sharp$ or GNA.
In light of these results we conclude that luminous GW Vir stars (i.e., those before the knee, as seen in Fig. \ref{FigHR} in the HR diagram offer a good testing ground for the occurrence of these mixing process.
With this in mind, we  chose GW Vir stars with the longest observed periods (PG 1159-035, NGC 1501, HS 2324, and NGC 246) and performed a residuals analysis of their periods in light of the predictions of our theoretical models. This qualitative comparison was inconclusive, with HS 2324 favoring models with smooth chemical profiles, while NGC 246 and NGC 1501 showed larger deviations from the mean period spacing, typical of our models with strong O-peaks in the outermost part of the core. Interestingly, we find that the prototype of the class, PG 1159-035, shows much larger mode trapping features than those observed in any of our models. We speculate that this could be related to the thinner envelopes (in comparison with canonical post-VLTP models) already hinted by \cite{2008ApJ...677L..35A} in light of the very large rates of period change displayed by this star. Besides being inconclusive, we believe our work shows that GW Vir stars offer a great opportunity to test chemically driven mixing processes on the TP-AGB phase.

Our results show that MLT$\sharp$ reproduces the predictions of the GNA theory with very good agreement, but through a significantly simpler and more numerically stable formulation. This makes it particularly suitable for stellar evolution codes where the implementation of GNA would be computationally expensive or unstable. Since MLT$\sharp$ modifies standard MLT only by incorporating the chemical gradient into the equations, it retains both the clarity of the physical interpretation and the flexibility of existing MLT-based frameworks. We conclude that MLT$\sharp$ is a practical and robust tool for modeling chemically driven convection and thermohaline mixing. This approach can be readily adopted in future evolutionary and pulsational studies where double-diffusive effects are relevant.

\begin{acknowledgements}
      Part of this work was supported by PIP-2971 from CONICET (Argentina) and by PICT 2020-03316 from Agencia I+D+i (Argentina). We thank the Asociación
Argentina de Astronomía for supporting the publication costs of this article.
\end{acknowledgements}

%
%
\bibliographystyle{aa}
\bibliography{bibliografia}

@ARTICLE{1991ApJ...367..601B,
       author = {{Brassard}, P. and {Fontaine}, G. and {Wesemael}, F. and {Kawaler}, S.~D. and {Tassoul}, M.},
        title = "{Adiabatic Properties of Pulsating DA White Dwarfs. I. The Treatment of the Brunt-Vaeisaelae Frequency and the Region of Period Formation}",
      journal = {\apj},
         year = 1991,
        month = feb,
       volume = {367},
        pages = {601},
          doi = {10.1086/169655},
       adsurl = {https://ui.adsabs.harvard.edu/abs/1991ApJ...367..601B}
}

@ARTICLE{2008ApJ...677L..35A,
       author = {{Althaus}, L.~G. and {C{\'o}rsico}, A.~H. and {Miller Bertolami}, M.~M. and {Garc{\'\i}a-Berro}, E. and {Kepler}, S.~O.},
        title = "{Evidence of Thin Helium Envelopes in PG 1159 Stars}",
      journal = {\apjl},
         year = 2008,
        month = apr,
       volume = {677},
       number = {1},
        pages = {L35},
          doi = {10.1086/587739},
archivePrefix = {arXiv},
       eprint = {0802.3363},
 primaryClass = {astro-ph},
       adsurl = {https://ui.adsabs.harvard.edu/abs/2008ApJ...677L..35A}
}

@ARTICLE{2007A&A...475..619C,
       author = {{C{\'o}rsico}, A.~H. and {Miller Bertolami}, M.~M. and {Althaus}, L.~G. and {Vauclair}, G. and {Werner}, K.},
        title = "{Asteroseismological constraints on the coolest GW Virginis variable star (PG 1159-type) <ASTROBJ>PG 0122+200</ASTROBJ>}",
      journal = {\aap},
         year = 2007,
        month = nov,
       volume = {475},
       number = {2},
        pages = {619-627},
          doi = {10.1051/0004-6361:20078145},
archivePrefix = {arXiv},
       eprint = {0709.0280},
 primaryClass = {astro-ph},
       adsurl = {https://ui.adsabs.harvard.edu/abs/2007A&A...475..619C}
}

@ARTICLE{2022MNRAS.511L..66W,
       author = {{Werner}, Klaus and {Reindl}, Nicole and {Geier}, Stephan and {Pritzkuleit}, Max},
        title = "{Discovery of hot subdwarfs covered with helium-burning ash}",
      journal = {\mnras},
         year = 2022,
        month = mar,
       volume = {511},
       number = {1},
        pages = {L66-L71},
          doi = {10.1093/mnrasl/slac005},
archivePrefix = {arXiv},
       eprint = {2202.05633},
 primaryClass = {astro-ph.SR},
       adsurl = {https://ui.adsabs.harvard.edu/abs/2022MNRAS.511L..66W}
}

@ARTICLE{2022MNRAS.511L..60M,
       author = {{Miller Bertolami}, M.~M. and {Battich}, T. and {C{\'o}rsico}, A.~H. and {Althaus}, L.~G. and {Wachlin}, F.~C.},
        title = "{An evolutionary channel for CO-rich and pulsating He-rich subdwarfs}",
      journal = {\mnras},
         year = 2022,
        month = mar,
       volume = {511},
       number = {1},
        pages = {L60-L65},
          doi = {10.1093/mnrasl/slab134},
archivePrefix = {arXiv},
       eprint = {2202.05635},
 primaryClass = {astro-ph.SR},
       adsurl = {https://ui.adsabs.harvard.edu/abs/2022MNRAS.511L..60M}
}

@ARTICLE{2006PASP..118..183W,
       author = {{Werner}, Klaus and {Herwig}, Falk},
        title = "{The Elemental Abundances in Bare Planetary Nebula Central Stars and the Shell Burning in AGB Stars}",
      journal = {\pasp},
         year = 2006,
        month = feb,
       volume = {118},
       number = {840},
        pages = {183-204},
          doi = {10.1086/500443},
archivePrefix = {arXiv},
       eprint = {astro-ph/0512320},
 primaryClass = {astro-ph},
       adsurl = {https://ui.adsabs.harvard.edu/abs/2006PASP..118..183W}
}

@ARTICLE{2000A&A...360..952H,
       author = {{Herwig}, F.},
        title = "{The evolution of AGB stars with convective overshoot}",
      journal = {\aap},
         year = 2000,
        month = aug,
       volume = {360},
        pages = {952-968},
          doi = {10.48550/arXiv.astro-ph/0007139},
archivePrefix = {arXiv},
       eprint = {astro-ph/0007139},
 primaryClass = {astro-ph},
       adsurl = {https://ui.adsabs.harvard.edu/abs/2000A&A...360..952H}
}

@ARTICLE{2006NuPhA.777..311S,
       author = {{Straniero}, Oscar and {Gallino}, Roberto and {Cristallo}, Sergio},
        title = "{s process in low-mass asymptotic giant branch stars}",
      journal = {\nphysa},
         year = 2006,
        month = oct,
       volume = {777},
        pages = {311-339},
          doi = {10.1016/j.nuclphysa.2005.01.011},
archivePrefix = {arXiv},
       eprint = {astro-ph/0501405},
 primaryClass = {astro-ph},
       adsurl = {https://ui.adsabs.harvard.edu/abs/2006NuPhA.777..311S}
}

@ARTICLE{2020MNRAS.493.4748W,
       author = {{Wagstaff}, G. and {Miller Bertolami}, M.~M. and {Weiss}, A.},
        title = "{Impact of convective boundary mixing on the TP-AGB}",
      journal = {\mnras},
         year = 2020,
        month = apr,
       volume = {493},
       number = {4},
        pages = {4748-4762},
          doi = {10.1093/mnras/staa362},
archivePrefix = {arXiv},
       eprint = {2002.01860},
 primaryClass = {astro-ph.SR},
       adsurl = {https://ui.adsabs.harvard.edu/abs/2020MNRAS.493.4748W}
}

@INPROCEEDINGS{2015ASPC..493...83M,
       author = {{Miller Bertolami}, M.~M.},
        title = "{Post-Asymptotic Giant Branch Evolution of Low- and Intermediate-Mass Stars. Preliminary Results}",
    booktitle = {19th European Workshop on White Dwarfs},
         year = 2015,
       editor = {{Dufour}, P. and {Bergeron}, P. and {Fontaine}, G.},
       series = {Astronomical Society of the Pacific Conference Series},
       volume = {493},
        month = jun,
        pages = {83},
       adsurl = {https://ui.adsabs.harvard.edu/abs/2015ASPC..493...83M}
}

@ARTICLE{2024Galax..12...83M,
       author = {{Miller Bertolami}, Marcelo M.},
        title = "{Primer on Formation and Evolution of Hydrogen-Deficient Central Stars of Planetary Nebul{\ae} and Related Objects}",
      journal = {Galaxies},
         year = 2024,
        month = nov,
       volume = {12},
       number = {6},
          eid = {83},
        pages = {83},
          doi = {10.3390/galaxies12060083},
archivePrefix = {arXiv},
       eprint = {2411.18035},
 primaryClass = {astro-ph.SR},
       adsurl = {https://ui.adsabs.harvard.edu/abs/2024Galax..12...83M}
}

@ARTICLE{2021FrASS...7...95X,
       author = {{Xiong}, Da-run},
        title = "{Convection Theory and Relevant Problems in Stellar Structure, Evolution and Pulsation Stability Part Ⅰ. Convection Theory and Structure of convection zone and stellar evolution}",
      journal = {Frontiers in Astronomy and Space Sciences},
         year = 2021,
        month = may,
       volume = {7},
          eid = {95},
        pages = {95},
          doi = {10.3389/fspas.2020.438864},
       adsurl = {https://ui.adsabs.harvard.edu/abs/2021FrASS...7...95X}
}

@ARTICLE{2015ApJ...809...30A,
       author = {{Arnett}, W. David and {Meakin}, Casey and {Viallet}, Maxime and {Campbell}, Simon W. and {Lattanzio}, John C. and {Moc{\'a}k}, Miroslav},
        title = "{Beyond Mixing-length Theory: A Step Toward 321D}",
      journal = {\apj},
         year = 2015,
        month = aug,
       volume = {809},
       number = {1},
          eid = {30},
        pages = {30},
          doi = {10.1088/0004-637X/809/1/30},
archivePrefix = {arXiv},
       eprint = {1503.00342},
 primaryClass = {astro-ph.SR},
       adsurl = {https://ui.adsabs.harvard.edu/abs/2015ApJ...809...30A}
}

@ARTICLE{1953ZA.....32..135V,
       author = {{Vitense}, E.},
        title = "{Die Wasserstoffkonvektionszone der Sonne. Mit 11 Textabbildungen}",
      journal = {\zap},
         year = 1953,
        month = jan,
       volume = {32},
        pages = {135},
       adsurl = {https://ui.adsabs.harvard.edu/abs/1953ZA.....32..135V}
}

@ARTICLE{1966PASJ...18..374K,
       author = {{Kato}, S.},
        title = "{Overstable Convection in a Medium Stratified in Mean Molecular Weight}",
      journal = {\pasj},
         year = 1966,
        month = jan,
       volume = {18},
        pages = {374},
       adsurl = {https://ui.adsabs.harvard.edu/abs/1966PASJ...18..374K}
}

@ARTICLE{2023A&A...680A.101S,
       author = {{Sibony}, Y. and {Georgy}, C. and {Ekstr{\"o}m}, S. and {Meynet}, G.},
        title = "{The impact of convective criteria on the properties of massive stars}",
      journal = {\aap},
         year = 2023,
        month = dec,
       volume = {680},
          eid = {A101},
        pages = {A101},
          doi = {10.1051/0004-6361/202346638},
archivePrefix = {arXiv},
       eprint = {2310.18139},
 primaryClass = {astro-ph.SR},
       adsurl = {https://ui.adsabs.harvard.edu/abs/2023A&A...680A.101S}
}

@ARTICLE{2022ApJ...926..169A,
       author = {{Anders}, Evan H. and {Jermyn}, Adam S. and {Lecoanet}, Daniel and {Brown}, Benjamin P.},
        title = "{Stellar Convective Penetration: Parameterized Theory and Dynamical Simulations}",
      journal = {\apj},
         year = 2022,
        month = feb,
       volume = {926},
       number = {2},
          eid = {169},
        pages = {169},
          doi = {10.3847/1538-4357/ac408d},
archivePrefix = {arXiv},
       eprint = {2110.11356},
 primaryClass = {astro-ph.SR},
       adsurl = {https://ui.adsabs.harvard.edu/abs/2022ApJ...926..169A}
}

@ARTICLE{2017RSOS....470192S,
       author = {{Salaris}, Maurizio and {Cassisi}, Santi},
        title = "{Chemical element transport in stellar evolution models}",
      journal = {Royal Society Open Science},
         year = 2017,
        month = aug,
       volume = {4},
       number = {8},
          eid = {170192},
        pages = {170192},
          doi = {10.1098/rsos.170192},
archivePrefix = {arXiv},
       eprint = {1707.07454},
 primaryClass = {astro-ph.SR},
       adsurl = {https://ui.adsabs.harvard.edu/abs/2017RSOS....470192S}
}

@ARTICLE{1932Biermann,
       author = {{Biermann}, L.},
        title = "{Untersuchungen {\"u}ber den inneren Aufbau der Sterne. IV. Konvektionszonen im Innern der Sterne. (Ver{\"o}ffentlichungen der Universit{\"a}ts-Sternwarte G{\"o}ttingen, Nr. 27. ) Mit 5 Abbildungen.}",
      journal = {\zap},
         year = 1932,
        month = jan,
       volume = {5},
        pages = {117},
       adsurl = {https://ui.adsabs.harvard.edu/abs/1932ZA......5..117B}
}

@ARTICLE{1925Prandtl,
       author = {{Prandtl}, L.},
        title = "{7. Bericht {\"u}ber Untersuchungen zur ausgebildeten Turbulenz}",
      journal = {Zeitschrift Angewandte Mathematik und Mechanik},
         year = 1925,
        month = jan,
       volume = {5},
       number = {2},
        pages = {136-139},
          doi = {10.1002/zamm.19250050212},
       adsurl = {https://ui.adsabs.harvard.edu/abs/1925ZaMM....5..136P}
}

@BOOK{Kipphenhahn2013,
       author = {{Kippenhahn}, Rudolf and {Weigert}, Alfred and {Weiss}, Achim},
        title = "{Stellar Structure and Evolution}",
         year = 2013,
          doi = {10.1007/978-3-642-30304-3},
       adsurl = {https://ui.adsabs.harvard.edu/abs/2013sse..book.....K}
}

@ARTICLE{2010A&ARv..18..471A,
       author = {{Althaus}, Leandro G. and {C{\'o}rsico}, Alejandro H. and {Isern}, Jordi and {Garc{\'\i}a-Berro}, Enrique},
        title = "{Evolutionary and pulsational properties of white dwarf stars}",
      journal = {\aapr},
         year = 2010,
        month = oct,
       volume = {18},
       number = {4},
        pages = {471-566},
          doi = {10.1007/s00159-010-0033-1},
archivePrefix = {arXiv},
       eprint = {1007.2659},
 primaryClass = {astro-ph.SR},
       adsurl = {https://ui.adsabs.harvard.edu/abs/2010A&ARv..18..471A}
}

@ARTICLE{2019A&ARv..27....7C,
       author = {{C{\'o}rsico}, Alejandro H. and {Althaus}, Leandro G. and {Miller Bertolami}, Marcelo M. and {Kepler}, S.~O.},
        title = "{Pulsating white dwarfs: new insights}",
      journal = {\aapr},
         year = 2019,
        month = sep,
       volume = {27},
       number = {1},
          eid = {7},
        pages = {7},
          doi = {10.1007/s00159-019-0118-4},
archivePrefix = {arXiv},
       eprint = {1907.00115},
 primaryClass = {astro-ph.SR},
       adsurl = {https://ui.adsabs.harvard.edu/abs/2019A&ARv..27....7C}
}

@ARTICLE{1958ZA.....46..108B,
       author = {{B{\"o}hm-Vitense}, E.},
        title = "{{\"U}ber die Wasserstoffkonvektionszone in Sternen verschiedener Effektivtemperaturen und Leuchtkr{\"a}fte. Mit 5 Textabbildungen}",
      journal = {\zap},
         year = 1958,
        month = jan,
       volume = {46},
        pages = {108},
       adsurl = {https://ui.adsabs.harvard.edu/abs/1958ZA.....46..108B}
}

@ARTICLE{1993ApJ...407..284G,
       author = {{Grossman}, Scott A. and {Narayan}, Ramesh and {Arnett}, David},
        title = "{A Theory of Nonlocal Mixing-Length Convection. I. The Moment Formalism}",
      journal = {\apj},
         year = 1993,
        month = apr,
       volume = {407},
        pages = {284},
          doi = {10.1086/172513},
       adsurl = {https://ui.adsabs.harvard.edu/abs/1993ApJ...407..284G}
}

@ARTICLE{1996MNRAS.279..305G,
       author = {{Grossman}, Scott A.},
        title = "{A theory of non-local mixing-length convection - III. Comparing theory and numerical experiment}",
      journal = {\mnras},
         year = 1996,
        month = mar,
       volume = {279},
       number = {2},
        pages = {305-336},
          doi = {10.1093/mnras/279.2.305},
archivePrefix = {arXiv},
       eprint = {astro-ph/9509054},
 primaryClass = {astro-ph},
       adsurl = {https://ui.adsabs.harvard.edu/abs/1996MNRAS.279..305G}
}

@ARTICLE{2011A&A...533A.139W,
       author = {{Wachlin}, F.~C. and {Miller Bertolami}, M.~M. and {Althaus}, L.~G.},
        title = "{Thermohaline mixing and the photospheric composition of low-mass giant stars}",
      journal = {\aap},
         year = 2011,
        month = sep,
       volume = {533},
          eid = {A139},
        pages = {A139},
          doi = {10.1051/0004-6361/201117029},
archivePrefix = {arXiv},
       eprint = {1104.0832},
 primaryClass = {astro-ph.SR},
       adsurl = {https://ui.adsabs.harvard.edu/abs/2011A&A...533A.139W}
}

@ARTICLE{1980A&A....91..175K,
       author = {{Kippenhahn}, R. and {Ruschenplatt}, G. and {Thomas}, H. -C.},
        title = "{The time scale of thermohaline mixing in stars}",
      journal = {\aap},
         year = 1980,
        month = nov,
       volume = {91},
       number = {1-2},
        pages = {175-180},
       adsurl = {https://ui.adsabs.harvard.edu/abs/1980A&A....91..175K}
}

@ARTICLE{2005ARA&A..43..435H,
       author = {{Herwig}, Falk},
        title = "{Evolution of Asymptotic Giant Branch Stars}",
      journal = {\araa},
         year = 2005,
        month = sep,
       volume = {43},
       number = {1},
        pages = {435-479},
          doi = {10.1146/annurev.astro.43.072103.150600},
       adsurl = {https://ui.adsabs.harvard.edu/abs/2005ARA&A..43..435H}
}

@ARTICLE{2016A&A...588A..25M,
       author = {{Miller Bertolami}, Marcelo Miguel},
        title = "{New models for the evolution of post-asymptotic giant branch stars and central stars of planetary nebulae}",
      journal = {\aap},
         year = 2016,
        month = apr,
       volume = {588},
          eid = {A25},
        pages = {A25},
          doi = {10.1051/0004-6361/201526577},
archivePrefix = {arXiv},
       eprint = {1410.1679},
 primaryClass = {astro-ph.SR},
       adsurl = {https://ui.adsabs.harvard.edu/abs/2016A&A...588A..25M}
}

@ARTICLE{2020A&A...633A..20A,
       author = {{Althaus}, Leandro G. and {C{\'o}rsico}, Alejandro H. and {Uzundag}, Murat and {Vu{\v{c}}kovi{\'c}}, Maja and {Baran}, Andrzej S. and {Bell}, Keaton J. and {Camisassa}, Mar{\'\i}a E. and {Calcaferro}, Leila M. and {De Ger{\'o}nimo}, Francisco C. and {Kepler}, Souza Oliveira and {Silvotti}, Roberto},
        title = "{About the existence of warm H-rich pulsating white dwarfs}",
      journal = {\aap},
         year = 2020,
        month = jan,
       volume = {633},
          eid = {A20},
        pages = {A20},
          doi = {10.1051/0004-6361/201936346},
archivePrefix = {arXiv},
       eprint = {1911.02442},
 primaryClass = {astro-ph.SR},
       adsurl = {https://ui.adsabs.harvard.edu/abs/2020A&A...633A..20A}
}

@ARTICLE{1990ApJS...72..335T,
       author = {{Tassoul}, M. and {Fontaine}, G. and {Winget}, D.~E.},
        title = "{Evolutionary Models for Pulsation Studies of White Dwarfs}",
      journal = {\apjs},
         year = 1990,
        month = feb,
       volume = {72},
        pages = {335},
          doi = {10.1086/191420},
       adsurl = {https://ui.adsabs.harvard.edu/abs/1990ApJS...72..335T}
}

@ARTICLE{2006A&A...454..863C,
       author = {{C{\'o}rsico}, A.~H. and {Althaus}, L.~G.},
        title = "{Asteroseismic inferences on GW Virginis variable stars in the frame of new PG 1159 evolutionary models}",
      journal = {\aap},
         year = 2006,
        month = aug,
       volume = {454},
       number = {3},
        pages = {863-881},
          doi = {10.1051/0004-6361:20054199},
archivePrefix = {arXiv},
       eprint = {astro-ph/0603736},
 primaryClass = {astro-ph},
       adsurl = {https://ui.adsabs.harvard.edu/abs/2006A&A...454..863C}
}

@ARTICLE{2024ApJ...969...10C,
       author = {{Castro-Tapia}, Matias and {Cumming}, Andrew and {Fuentes}, J.~R.},
        title = "{Fast and Slow Crystallization-driven Convection in White Dwarfs}",
      journal = {\apj},
         year = 2024,
        month = jul,
       volume = {969},
       number = {1},
          eid = {10},
        pages = {10},
          doi = {10.3847/1538-4357/ad4152},
archivePrefix = {arXiv},
       eprint = {2402.01947},
 primaryClass = {astro-ph.SR},
       adsurl = {https://ui.adsabs.harvard.edu/abs/2024ApJ...969...10C}
}

@ARTICLE{1983A&A...126..207L,
       author = {{Langer}, N. and {Fricke}, K.~J. and {Sugimoto}, D.},
        title = "{Semiconvective diffusion and energy transport}",
      journal = {\aap},
         year = 1983,
        month = sep,
       volume = {126},
       number = {1},
        pages = {207},
       adsurl = {https://ui.adsabs.harvard.edu/abs/1983A&A...126..207L}
}

@ARTICLE{1996MNRAS.283.1165G,
       author = {{Grossman}, Scott A. and {Taam}, Ronald E.},
        title = "{Double-Diffusive Mixing-Length Theory, Semiconvection and Massive Star Evolution}",
      journal = {\mnras},
         year = 1996,
        month = dec,
       volume = {283},
       number = {4},
        pages = {1165-1178},
          doi = {10.1093/mnras/283.4.1165},
archivePrefix = {arXiv},
       eprint = {astro-ph/9608137},
 primaryClass = {astro-ph},
       adsurl = {https://ui.adsabs.harvard.edu/abs/1996MNRAS.283.1165G}
}

@INPROCEEDINGS{2020mdps.conf...13G,
       author = {{Garaud}, Pascale},
        title = "{Double-diffusive processes in stellar astrophysics}",
    booktitle = {Multi-Dimensional Processes In Stellar Physics},
         year = 2020,
       editor = {{Rieutord}, Michel and {Baraffe}, Isabelle and {Lebreton}, Yveline},
        month = jan,
        pages = {13},
       adsurl = {https://ui.adsabs.harvard.edu/abs/2020mdps.conf...13G}
}

@ARTICLE{1988Ap&SS.150..115U,
       author = {{Umezu}, Minoru and {Nakakita}, Tomofumi},
        title = "{The Local Mixing-Length Theory with Convective Helium Flux}",
      journal = {\apss},
         year = 1988,
        month = dec,
       volume = {150},
       number = {1},
        pages = {115-147},
          doi = {10.1007/BF00714158},
       adsurl = {https://ui.adsabs.harvard.edu/abs/1988Ap&SS.150..115U}
}

@ARTICLE{2021A&A...645A.117C,
       author = {{C{\'o}rsico}, A.~H. and {Uzundag}, M. and {Kepler}, S.~O. and {Althaus}, L.~G. and {Silvotti}, R. and {Baran}, A.~S. and {Vu{\v{c}}kovi{\'c}}, M. and {Werner}, K. and {Bell}, K.~J. and {Higgins}, M.},
        title = "{Pulsating hydrogen-deficient white dwarfs and pre-white dwarfs observed with TESS. I. Asteroseismology of the GW Vir stars RX J2117+3412, HS 2324+3944, NGC 6905, NGC 1501, NGC 2371, and K 1-16}",
      journal = {\aap},
         year = 2021,
        month = jan,
       volume = {645},
          eid = {A117},
        pages = {A117},
          doi = {10.1051/0004-6361/202039202},
archivePrefix = {arXiv},
       eprint = {2011.03629},
 primaryClass = {astro-ph.SR},
       adsurl = {https://ui.adsabs.harvard.edu/abs/2021A&A...645A.117C}
}

@ARTICLE{2022ApJ...936..187O,
       author = {{Oliveira da Rosa}, Gabriela and {Kepler}, S.~O. and {C{\'o}rsico}, Alejandro H. and {Costa}, J.~E.~S. and {Hermes}, J.~J. and {Kawaler}, S.~D. and {Bell}, Keaton J. and {Montgomery}, M.~H. and {Provencal}, J.~L. and {Winget}, D.~E. and {Handler}, G. and {Dunlap}, Bart and {Clemens}, J.~C. and {Uzundag}, Murat},
        title = "{Kepler and TESS Observations of PG 1159-035}",
      journal = {\apj},
         year = 2022,
        month = sep,
       volume = {936},
       number = {2},
          eid = {187},
        pages = {187},
          doi = {10.3847/1538-4357/ac8871},
archivePrefix = {arXiv},
       eprint = {2208.04791},
 primaryClass = {astro-ph.SR},
       adsurl = {https://ui.adsabs.harvard.edu/abs/2022ApJ...936..187O}
}

@ARTICLE{1906WisGo.195...41S,
       author = {{Schwarzschild}, K.},
        title = "{On the equilibrium of the Sun's atmosphere}",
      journal = {Nachrichten von der K{\"o}niglichen Gesellschaft der Wissenschaften zu G{\"o}ttingen. Math.-phys. Klasse},
         year = 1906,
        month = jan,
       volume = {195},
        pages = {41-53},
       adsurl = {https://ui.adsabs.harvard.edu/abs/1906WisGo.195...41S}
}

@ARTICLE{1947ApJ...105..305L,
       author = {{Ledoux}, P.},
        title = "{Stellar Models with Convection and with Discontinuity of the Mean Molecular Weight}",
      journal = {\apj},
         year = 1947,
        month = mar,
       volume = {105},
        pages = {305},
          doi = {10.1086/144905},
       adsurl = {https://ui.adsabs.harvard.edu/abs/1947ApJ...105..305L}
}

@ARTICLE{1959PThPh..22..830S,
       author = {{Sakashita}, S. and {Hayashi}, C.},
        title = "{Internal Structure and Evolution of Very Massive Stars}",
      journal = {Progress of Theoretical Physics},
         year = 1959,
        month = dec,
       volume = {22},
       number = {6},
        pages = {830-834},
          doi = {10.1143/PTP.22.830},
       adsurl = {https://ui.adsabs.harvard.edu/abs/1959PThPh..22..830S}
}

@INPROCEEDINGS{1979wdvd.coll..377M,
       author = {{McGraw}, J.~T. and {Starrfield}, S.~G. and {Liebert}, J. and {Green}, R.},
        title = "{PG 1159-035: a New, Hot, Non-Da Pulsating Degenerate}",
    booktitle = {IAU Colloq. 53: White Dwarfs and Variable Degenerate Stars},
         year = 1979,
       editor = {{van Horn}, H.~M. and {Weidemann}, V. and {Savedoff}, M.~P.},
        month = jan,
        pages = {377},
       adsurl = {https://ui.adsabs.harvard.edu/abs/1979wdvd.coll..377M}
}

@ARTICLE{2024A&A...686A.140C,
       author = {{Calcaferro}, Leila M. and {Sowicka}, Paulina and {Uzundag}, Murat and {C{\'o}rsico}, Alejandro H. and {Kepler}, Souza O. and {Bell}, Keaton J. and {Althaus}, Leandro G. and {Handler}, Gerald and {Kawaler}, Steven D. and {Werner}, Klaus},
        title = "{Pulsating hydrogen-deficient white dwarfs and pre-white dwarfs observed with TESS. VI. Asteroseismology of the GW Vir-type central star of the Planetary Nebula NGC 246}",
      journal = {\aap},
         year = 2024,
        month = jun,
       volume = {686},
          eid = {A140},
        pages = {A140},
          doi = {10.1051/0004-6361/202349103},
archivePrefix = {arXiv},
       eprint = {2402.16642},
 primaryClass = {astro-ph.SR},
       adsurl = {https://ui.adsabs.harvard.edu/abs/2024A&A...686A.140C}
}

@ARTICLE{2024A&A...691A.194C,
       author = {{Calcaferro}, Leila M. and {C{\'o}rsico}, Alejandro H. and {Uzundag}, Murat and {Althaus}, Leandro G. and {Kepler}, S.~O. and {Werner}, Klaus},
        title = "{An analysis of spectroscopic, seismological, astrometric, and photometric masses of pulsating white dwarf stars}",
      journal = {\aap},
         year = 2024,
        month = nov,
       volume = {691},
          eid = {A194},
        pages = {A194},
          doi = {10.1051/0004-6361/202450582},
archivePrefix = {arXiv},
       eprint = {2409.03896},
 primaryClass = {astro-ph.SR},
       adsurl = {https://ui.adsabs.harvard.edu/abs/2024A&A...691A.194C}
}

@ARTICLE{2006A&A...458..259C,
       author = {{C{\'o}rsico}, A.~H. and {Althaus}, L.~G. and {Miller Bertolami}, M.~M.},
        title = "{New nonadiabatic pulsation computations on full PG 1159 evolutionary models: the theoretical GW Virginis instability strip revisited}",
      journal = {\aap},
         year = 2006,
        month = oct,
       volume = {458},
       number = {1},
        pages = {259-267},
          doi = {10.1051/0004-6361:20065423},
archivePrefix = {arXiv},
       eprint = {astro-ph/0607012},
 primaryClass = {astro-ph},
       adsurl = {https://ui.adsabs.harvard.edu/abs/2006A&A...458..259C}
}

@INPROCEEDINGS{2005ASPC..334..553M,
       author = {{Montgomery}, M.~H.},
        title = "{The Vibrating String Analogy and Mode Trapping}",
    booktitle = {14th European Workshop on White Dwarfs},
         year = 2005,
       editor = {{Koester}, D. and {Moehler}, S.},
       series = {Astronomical Society of the Pacific Conference Series},
       volume = {334},
        month = jul,
        pages = {553},
       adsurl = {https://ui.adsabs.harvard.edu/abs/2005ASPC..334..553M}
}

\begin{appendix} 
\section{Mathematical derivation of MLT$\sharp$ equations} \label{AppendixA}

In this section, we derive the equations that rule the convective motions in MLT$\sharp$. We will follow an almost identical approach as in Chapter 7 of \cite{Kipphenhahn2013}, indicating where the differences arise.

Considering the sum of radiative flux $F_\text{rad}$ and convective flux, $F_\text{con}$, we obtain the radiative gradient $\nabla_\text{rad}$ needed to transport the whole flux by radiation,
\begin{equation}\label{A1}
    F_\text{rad}+F_\text{con}=\frac{4acGT^4m}{3\kappa P r^2}\nabla_\text{rad}
.\end{equation}

Considering that part of the flux is transported by convection, then $F_\text{rad}$ is related to the actual temperature gradient $\nabla$ by
\begin{equation}\label{A2}
    F_\text{rad}=\frac{4acGT^4m}{3\kappa P r^2}\nabla.
\end{equation}

On the other hand, the local convective flux of energy given by an element moving with a velocity $v$ and in balance of pressure ($DP=0$) and excess temperature $DT$\footnote{We recall that for a given quantity $X$, its difference $DX$ is $DX=X_\text{e} - X_\text{s}$. However, we note that in Eq. \ref{A6} the difference $D\mu$ is only given by the chemical gradient of the background, since we are assuming that the convective element suffers no changes in its internal chemical composition.} is given by
\begin{equation} \label{A3}
    F_\text{con}=\rho v c_P DT.
\end{equation}
The element will move over a distance $l_m$, the \textit{mixing length}, until it mixes with the surrounding. Since the elements will start their motions at different distances relative to a given place (from $0$ to $l_m$), they will have different velocities $v$ and excess temperatures $DT$. We can consider an ``average'' element that has moved $l_m/2$ and then we have
\begin{equation} \label{A4}
    \frac{DT}{T} = \frac{1}{T}\frac{\partial DT}{\partial r} \frac{l_m}{2} = (\nabla - \nabla_\text{e}) \frac{l_m}{2H_P} .
\end{equation}

This is the point where the differences between the standard treatment of MLT and MLT$\sharp$ arise. Considering the density difference without neglecting the impact of the chemical instabilities, we have
\begin{equation}
     \frac{D\rho}{\rho} = -\delta \frac{DT}{T} + \varphi \frac{D\mu}{\mu},
\end{equation}
with
\begin{equation} \label{A6}
    \frac{D\mu}{\mu}=\nabla_\mu \frac{l_m}{2H_P}
\end{equation}
and the radial buoyancy force per unit mass is
\begin{equation}
    k_r = -g \frac{D\rho}{\rho}.
\end{equation}
Considering that, on average, half of this value acts on its motion, $l_m/2,  $ giving \begin{equation}
    \frac{1}{2}k_r\frac{l_m}{2} = g\delta (\nabla - \nabla_\text{e} )\frac{l_m^2}{8H_P} - \varphi \nabla_\mu \frac{l_m}{4} .
\end{equation}
Considering that half of the work goes into the kinetic energy, the average velocity of the elements will be given by

\begin{equation} \label{A9}
    v^2 = g\delta \bigg(\nabla - \nabla_\text{e} - \frac{\varphi}{\delta}\nabla_\mu\bigg)\frac{l_m^2}{8H_P}.
\end{equation}
Assuming that the chemical gradient $\nabla_\mu$ remains constant over the course of the timescale of the convective motion
we need to obtain only the value for $\nabla$ and $\nabla_\text{e}$. We will use the following relation, from \cite{Kipphenhahn2013}:
\begin{equation}
    \frac{\nabla_\text{e}-\nabla_\text{ad}}{\nabla-\nabla_\text{e}}=\frac{6acT^3}{\kappa\rho^2c_Pl_mv},
\end{equation}
where we assume the convective element to be a sphere of diameter $l_m$. This means that the mixing length is assumed to be the same as the size of the blob of matter.

Replacing \ref{A9} in the expression above, we can obtain the first of the equations that of MLT$\sharp$:
\begin{equation} \label{MLT1}
    \big(\nabla_\text{e}-\nabla_\text{ad}\big)\big(\nabla-\nabla_\text{e}-\frac{\varphi}{\delta}\nabla_\mu\big)^{1/2}=2U\big(\nabla-\nabla_\text{e}\big),
\end{equation}
with $U$ equally defined as in Section \ref{sec2}. 

 Next, in order to obtain the second equation, we can replace \ref{A4} and \ref{A9} in \ref{A3}, obtaining the following expression for the convective flux:
\begin{equation} \label{A13}
    F_\text{con}=\frac{\rho c_P\sqrt{g\delta}}{4\sqrt{2}}l_m^2 TH_P^{-3/2}\big(\nabla-\nabla_\text{e}-\frac{\varphi}{\delta}\nabla_\mu\big)^{1/2}(\nabla-\nabla_\text{e}).
\end{equation}

Also, from \ref{A1} and \ref{A2}, we have

\begin{equation}\label{A14}
    F_\text{con}=\frac{4acGT^4m}{3\kappa Pr^2}\big(\nabla_\text{rad}-\nabla\big).
\end{equation}
Finally, using $dP/dr=-Gm\rho/r^2$ and the definition for $H_P$, we can combine \ref{A13} and \ref{A14} in order to obtain the second equation of MLT$\sharp$, expressed as
\begin{equation} \label{MLT2}
    \big(\nabla-\nabla_\text{e}-\frac{\varphi}{\delta}\nabla_\mu\big)^{1/2}(\nabla-\nabla_\text{e})=\frac{8}{9}U\big(\nabla_\text{rad}-\nabla).
\end{equation}

Eqs. \ref{MLT1} and \ref{MLT2} can be solved numerically for obtaining $\nabla$ and $\nabla_\text{e}$. Replacing the gradients in \ref{A9} will give the convective velocity of the element.

\end{appendix}

\end{document}